\documentclass[prl, preprint, superscriptaddress, floatfix]{revtex4-1}

\usepackage{graphicx}
\usepackage{textgreek}
\usepackage{bm}
\usepackage[UKenglish]{babel}
\usepackage{xspace}
\usepackage{xr}
\usepackage{textcomp}
\usepackage{float}
\usepackage[final]{pdfpages}

\makeatletter
\AtBeginDocument{\let\LS@rot\@undefined}
\makeatother

\graphicspath{
{../../Figures/FigOverview/}
{../../Figures/FigHall/}
{../../Figures/FigNoise/}
{../../Figures/FigQH/}
{../../Figures/FigComparison/}
{../../Figures/Supp/FigDevices/}
{../../Figures/Supp/FigHallsupp/}
{../../Figures/Supp/FigInhomogeneity/}
{../../Figures/Supp/FigRTN/}
{../../Figures/Supp/FigTemp/}
} 
\newcommand{\perroothz}{~Hz$^{-1/2}$\xspace}
\newcommand{\densityunits}{~cm$^{-2}$\xspace}
\newcommand{\degC}{~$^{\circ}$C\xspace}
\newcommand{\micron}{~\textmu m\xspace}
\newcommand{\capacitanceunits}{~\textmu F\,cm$^{-2}$\xspace}

\newcommand{\textapprox}{$\sim$}

\let\oldsubsection\subsection
\renewcommand{\subsection}[1]{\oldsubsection{#1}}

\begin{document}


# Gate-Tunable Graphene Hall Sensors with High Magnetic Field Sensitivity

Brian T. Schaefer[1], Lei Wang[2], Alexander Jarjour[1], Kenji Watanabe[3], Takashi Taniguchi[3], Paul L. McEuen[1,2], and Katja C. Nowack[1,2*]

[1]Laboratory of Atomic and Solid-State Physics, Cornell University, Ithaca, NY 14853, USA
[2]Kavli Institute at Cornell for Nanoscale Science, Cornell University, Ithaca, NY 14853, USA
[3]National Institute for Materials Science, 1-1 Namiki, Tsukuba 305-0044, Japan

*To whom correspondence should be addressed: kcn34@cornell.edu

**Solid-state magnetic field sensors are important to both modern electronics and fundamental materials science. Many types of these sensors maintain high sensitivity only in a limited range of temperature and background magnetic field, but Hall-effect sensors are in principle able to operate over a broad range of these conditions. Here, we fabricate and characterize micrometer-scale graphene Hall sensors demonstrating high magnetic field sensitivity from liquid-helium to room temperature and in background magnetic field up to several Tesla. By tuning the charge carrier density with an electrostatic gate, we optimize the magnetic field sensitivity for different working conditions. From measurements of the Hall coefficient and the Hall voltage noise at 1 kHz, we estimate an optimum magnetic field sensitivity of 80 nT Hz$^{-1/2}$ at 4.2 K, 700 nT Hz$^{-1/2}$ at room temperature, and 3 µT Hz$^{-1/2}$ in 3 T background magnetic field at 4.2 K. Our devices perform competitively with the best existing Hall sensor technologies at room temperature, outperform any Hall sensors reported in the literature at 4.2 K, and demonstrate high sensitivity for the first time in a few Tesla applied magnetic field.**

Hall-effect sensors are attractive for a variety of applications ranging from position detection in robotics[1] and tracking nanoparticles in biological systems[2] to fundamental studies of magnetism[3] and superconductivity[4–6]. Combining a versatile fabrication process and straightforward measurement scheme, Hall sensors provide an accessible means of performing non-invasive measurements of small magnetic fields. In an ideal Hall-effect sensor, the deflection of electric current in an out-of-plane magnetic field $B$ produces a transverse (Hall) voltage response $V_H = BI/(ne)$, where $I$ is the bias current, $n$ is the two-dimensional charge carrier density, and $e$ is the electron charge[7,8]. This voltage response and the Hall voltage noise $S_V^{1/2}$ determine the magnetic field sensitivity $S_B^{1/2} = S_V^{1/2}/(\partial V_H/\partial B)$. The fundamental limit for the noise performance is thermal Johnson noise, which is proportional to the square root of the device resistance, and the total noise typically has contributions from instrumentation noise, flicker noise, and random telegraph noise[8–11].

The desired combination of a large voltage response and low Johnson noise suggest a material system with low carrier density and high carrier mobility. Whereas carrier mobility decreases at low carrier density in most semiconductor-based two-dimensional electron systems[12], in graphene the mobility is enhanced at low carrier density in the absence of long-range impurity scattering[13]. Recently, encapsulation in hexagonal boron nitride (hBN) has enabled access of this low-density, high-mobility regime[14]. Here, we fabricate Hall sensors from encapsulated graphene and exploit tuning the carrier density via electrostatic gating to maintain high sensitivity under different operating conditions, including for the first time in a high background magnetic field. We demonstrate that these graphene devices are, depending on the operating regime, comparable to or better than the most sensitive Hall sensors reported in the literature.

Figure 1 summarizes our main result. We compare the best magnetic field sensitivity $S_B^{1/2}$ for our devices (black markers) with measurements for leading Hall sensors in the literature. We include work that both reports noise spectra and specifies micrometer-scale device dimensions[2,8–10,15–20]. The quantity $S_B^{1/2}$ multiplied by the square root of the measurement bandwidth is the smallest detectable change in magnetic field, typically reported at kHz frequencies. We choose a reference frequency of 1 kHz to avoid contributions from random telegraph noise dominant at lower frequency (see Figure 4 and accompanying discussion). The noise performance of Hall sensors depends strongly on the material and size of the device. Previous work shows that $S_B^{1/2}$ for a given material and fabrication procedure scales approximately as $w^{-1}$ with the width $w$ of the active Hall sensor area[15,16]. Devices with similar performance fall along the dashed diagonal lines of constant $S_B^{1/2}w$ in Figure 1. The best-performing devices minimize $S_B^{1/2}w$, combining low noise and small size. We find that our devices fall into the bottom left corner, outperforming any other sensor reported in the literature at low temperature. They also perform competitively with the best sensors made from InSb[18] and hBN-encapsulated graphene[8] at room temperature. Furthermore, even the sensitivity in high applied magnetic field is still competitive with most leading sensors characterized at zero magnetic field. To our knowledge, no reports of operating a micrometer-scale Hall sensor with high sensitivity in large magnetic field has been reported in the literature so far.

Each of our devices (Figure 2a) is fabricated on a silicon substrate and consists of exfoliated monolayer graphene encapsulated with hBN gate dielectrics[14] and few-layer graphite (FLG) gate electrodes[21–23] assembled using a dry-transfer technique (see Supporting Information for fabrication details). The combination of low charged defect density in hBN and the ability of FLG to screen charged impurity disorder in the silicon substrate improves carrier mobility, reduces the charge inhomogeneity, and can reduce charge noise in graphene devices[21–25]. The top graphite gate tunes the carrier density in the active region of the device, while the silicon back gate dopes the contacts to high electron density, lowering the contact resistance and voltage noise (see Supporting Information, Figure S2). In principle only a top graphite gate is needed, but our best-performing device (D1 in Figure 1) includes a lower graphite gate as well (see Supporting Information, Figure S2 for details on additional devices).

We first evaluate the electronic quality of our devices at low background magnetic field and low temperature in a liquid-helium cryostat. We bias the device with a small ac current $I$ and measure the two-point ($V_{2p}$) and Hall ($V_H$) voltages using standard low-frequency lock-in techniques while applying top gate voltage $V_g$ to tune the carrier density (Figure 2a,b). From a series of gate sweeps at fixed magnetic field $B$ up to 100 mT (see Supporting Information, Figure S1), we determine the Hall coefficient $R_H = I^{-1} (\partial V_H / \partial B)_{B=0}$ and extract the carrier density $n = (e R_H)^{-1}$ (Figure 2c, upper panel). At gate voltages near the charge neutrality point (CNP), the coexistence of electrons and holes makes the Hall voltage nonlinear in magnetic field[26]. Elsewhere, $R_H \sim n^{-1} \sim V_g^{-1}$ assuming a simple capacitive coupling of the gate to the mobile carrier density[7,13]. Extrapolating the electron and hole densities to zero reveals that electrons and holes appear to reach charge neutrality at different $V_g$. This is consistent with contributions to the charging behavior of the graphene sheet from the quantum capacitance[13,27] and additional charge traps with non-constant capacitance[7], which become significant because of the large gate capacitance and small charge inhomogeneity in our devices. The maximum (minimum) value of $R_H$ for electron (hole) doping 240 kΩ/T (−340 kΩ/T) implies a smallest mobile carrier density $\delta n \sim 2.6 \times 10^9$ cm$^{-2}$ (−1.8 × 10$^9$ cm$^{-2}$) limited by intrinsic charge inhomogeneity. Moreover, the width of the peak in the two-point resistance $R_{2p} = V_{2p}/I$ (Figure 2c, lower panel) implies a charge inhomogeneity ~ 4 × 10$^9$ cm$^{-2}$. This low amount of charge inhomogeneity is consistent with the best reported devices using atomically smooth single-crystal graphite flakes as gate electrodes[22,23].

Next, we characterize the voltage response as a function of applied dc current bias up to 50 μA. The Hall voltage response to a small change in magnetic field $\delta B$ is $\delta V_H = I R_H \delta B$, suggesting that applying a larger bias current in principle proportionally increases the voltage signal. In practice, a large dc bias causes two changes in the transport characteristics of the devices (Figure 3): the peak $R_H$ decreases and the CNP gate voltage $V_g^0$ shifts. The direction of the shift depends on the polarity of the applied current. These changes are consistent with a potential gradient and resulting carrier density gradient across the device[27,28] (see Supporting Information, Figure S3). This

modifies the average $R_H$ within the Hall cross and limits its peak value. Despite the reduction in peak $R_H$, applying larger bias current still increases the absolute voltage sensitivity $IR_H = (\partial V_H/\partial B)_{B=0}$ (Figure 3b), giving a larger change in Hall voltage per unit change in magnetic field.

To determine the magnetic field sensitivity reported in Figure 1, we measure the noise performance of the devices alongside the voltage response. We measure fluctuations in the Hall voltage in real time (Figure 4a) and take the Fourier transform to arrive at the Hall voltage noise spectral density $S_V^{1/2}$ (Figure 4b) (see Supporting Information for details). At low bias, 60 Hz and preamplifier input noise dominate the $S_V^{1/2}$ spectrum (Figure 4c). The shape of the noise spectra at higher bias suggest the presence of both flicker noise[11,29] ($1/f$ noise; $S_V^{1/2} \sim f^{-1/2}$) and random telegraph noise[30] (RTN; $S_V^{1/2}$ constant at low frequency, $S_V^{1/2} \sim f^1$ at high frequency), as reported previously in micrometer-scale Hall sensors[9,10] and graphene-based devices[31]. Flicker noise originates most likely from random charging and discharging events of an ensemble of charge traps, while RTN is characteristic of a single charge trap more strongly coupled to the device[29]. These charging events can induce fluctuations in both the carrier mobility and carrier density which are prominent in graphene-based devices at low carrier density[9,11,29,32]. Charge fluctuations that modulate the contact resistance and defect states in the substrate or etched edges of the device can couple strongly into the voltage noise, especially near charge neutrality where charge fluctuations are poorly screened[11,32]. We find that the behavior of the RTN changes between successive cooldowns and under different conditions of current bias and gate voltage, suggesting that it arises here from a single charged impurity strongly coupled to the device (see Supporting Information, Figure S4).

Figure 4e summarizes the low-temperature gate dependence of $S_V^{1/2}$ at zero $B$ and corresponding magnetic field sensitivity $S_B^{1/2} = S_V^{1/2}/(\partial V_H/\partial B) = S_V^{1/2}/(IR_H)$ at 20 μA current bias and 1 kHz. At this frequency, we avoid RTN dominant at lower frequencies; at higher frequencies, the noise is limited by the instrumentation noise floor, making the gate voltage dependence less apparent. Figure 4f shows that a 20 μA bias current minimizes the magnetic field noise. At this intermediate bias current, the increase in the voltage signal above the instrumentation noise floor is favorable over the reduction of $R_H$ at large bias current. Notably, the minimum $S_B^{1/2}$ does not occur at the same value of $V_g$ at which $R_H$ peaks. This indicates that the optimum working point of the Hall probe balances tuning away from the CNP to reduce $S_V^{1/2}$ and tuning close to the CNP to increase $R_H$. The minimum value, $S_B^{1/2} \sim 80$ nT Hz$^{-1/2}$ at 1 kHz (lowermost point in Figure 1), is to our knowledge the smallest magnetic field noise ever reported in a micrometer-scale Hall sensor at 4.2 K. At room temperature, repeating the Hall coefficient and Hall voltage noise measurements (see Figure S5c,d in the Supporting Information) reveals that the magnetic field noise is somewhat larger, but still competitive with the best Hall sensors reported in the literature (see Figure 1).

Finally, we characterize the magnetic field sensitivity in a large magnetic field. To our knowledge this has not been reported for any leading micrometer-scale Hall sensors. In a large applied

magnetic field, the Hall resistance develops plateaus (Figure 5a) spaced by $\Delta(V_H/I)^{-1} = 4e^2/h$ as expected for monolayer graphene in the quantum Hall regime[13]. At 5 μA dc current bias, these plateaus onset at ~500 mT. The deviation of the resistance plateaus from precise quantization is caused by the large bias current and the wide, extended Hall voltage contacts in our device (Figure 2a), which mix a significant fraction of the longitudinal resistance into the Hall resistance[33]. The Hall coefficient $R_H = I^{-1}(\partial V_H/\partial B)$ (Figure 5b-d) now reaches local minima at values of $(B, V_g)$ corresponding to the resistance plateaus. At high magnetic field (Figure 5d), the resistance plateaus flatten, and $R_H$ drops completely to zero. Repeating measurements of the Hall voltage noise as described above, at 3 T we obtain $S_B^{1/2} \sim 3$ μT Hz$^{-1/2}$ at optimum carrier density tuning (Figure 5d, $V_g \sim 0.8$ V). The higher noise compared to measurements at zero field is a result of both the reduced $R_H$ and a general increase in voltage noise in large background magnetic field, which is correlated with large longitudinal magnetoresistance and may also be attributed to charge fluctuations between localized and extended quantum Hall states[34,35].

In summary, we show that hBN-encapsulated monolayer graphene combined with few-layer graphite gates is an excellent material system for micrometer-scale Hall sensors. Because our devices are gate-tunable, we can optimize the magnetic field sensitivity over a large range of both temperature and magnetic field. We anticipate that optimization of the measurement itself (e.g. through lower noise pre-amplification) will further improve the reported sensitivities in the future. In addition to enabling high sensitivity, the dry-transfer fabrication process offers the flexibility to fabricate Hall sensors directly on top of materials of interest[2,3,6]. In the future, incorporation of these devices in a scanning probe will enable the imaging of magnetic fields over a range of temperatures and magnetic fields not accessed with a single probe to date. This imaging technique will provide a new window into a range of condensed matter systems including unconventional superconductors across their magnetic field-temperature phase diagram, magnetic-field-tuned phases of matter, and electric currents in regimes of electronic transport that appear at high temperature and magnetic field.

## Acknowledgements

## Author contributions

B.T.S. fabricated the devices and performed the measurements, with support from L.W. and A.J. K.W. and T.T. synthesized the hexagonal boron nitride crystals. P.L.M. and K.C.N. supervised the project. B.T.S. and K.C.N. wrote the manuscript, with input from L.W. and P.L.M.

## Competing interests

The authors declare no competing financial interest.

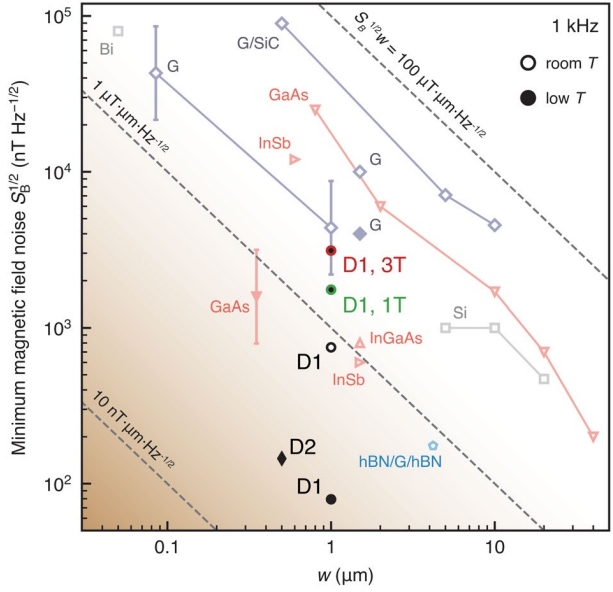

**Figure 1.** Magnetic field sensitivity $S_B^{1/2}$ at 1 kHz compared against the width $w$ of Hall sensors reported here and in the literature. The black markers show the best performance of our devices (D1, described above, and D2, a 500 nm device described in the Supporting Information). The other markers are estimates of the best performance of devices made from semiconductor- and graphene-based structures, including graphene grown by chemical vapor deposition (G), epitaxial graphene (G/SiC), and hBN-encapsulated exfoliated graphene (hBN/G/hBN). Markers connected by solid lines are from measurements on devices with the same material and fabrication process, showing an approximate $w^{-1}$ scaling (dashed line). Markers with error bars are extrapolated from measurements reported at lower frequencies, assuming the noise is dominated by flicker noise and scales as $f^{-\alpha}$ (error bars mark the range $0.4 < \alpha < 0.6$).

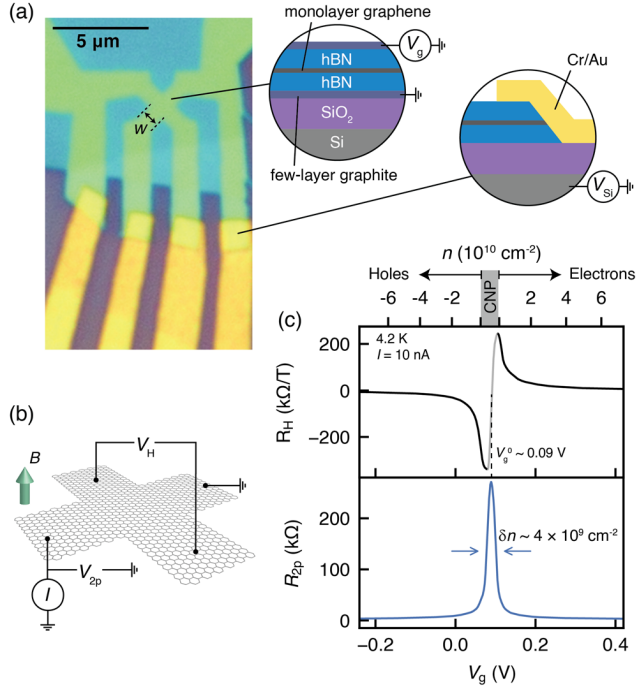

**Figure 2.** (a) Optical microscope image of a $w = 1$ μm graphene Hall sensor. Left cross-section: Hall cross layer structure consisting of monolayer graphene encapsulated with hexagonal boron nitride (hBN) and few-layer graphite. Right cross-section: edge contacts to graphene, doped to high electron density with the silicon gate ($V_{Si} = 40$ V). (b) Schematic of the measurement configuration: we measure the Hall voltage $V_H$ and two-point voltage $V_{2p}$ under bias current $I$ and out-of-plane magnetic field $B$. (c) Top gate voltage dependence of the Hall coefficient $R_H$ and two-point resistance $R_{2p}$ under small ac bias and using measurements up to $B = 100$ mT. The upper axis indicates the corresponding electron and hole densities.

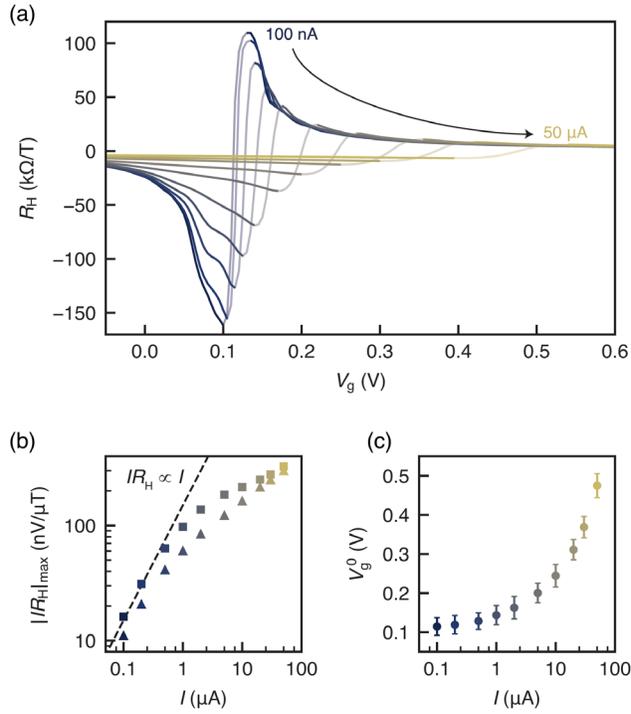

**Figure 3.** (a) Measurements of the Hall coefficient $R_H$ under dc current bias at 4.2 K. (b) Bias current dependence of the peak value of $IR_H$. (c) Bias current dependence of the charge neutrality point voltage $V_g^0$. Error bars represent the uncertainty in determining the point at which $R_H$ crosses zero.

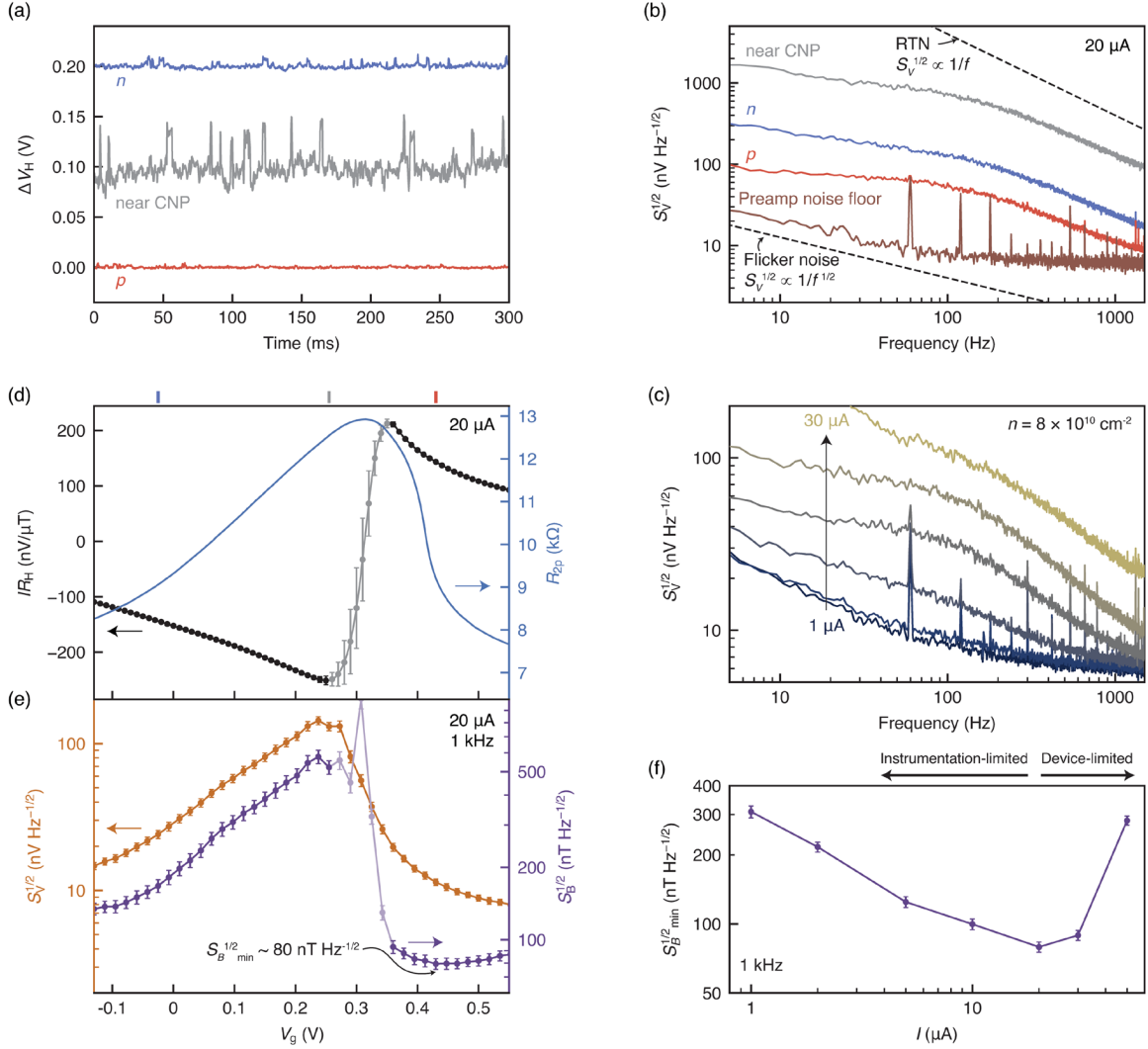

**Figure 4.** (a) Time traces of the Hall voltage (offset for clarity) and (b) Hall voltage noise spectral density $S_V^{1/2}$ for fixed bias current at 4.2 K. The three curves correspond to the gate voltages marked at the top of the upper panel of (d). Dashed lines in (b) follow the expected dependence of random telegraph noise (RTN) at high frequency ($f^{-1}$) and flicker noise ($f^{-1/2}$). (c) $S_V^{1/2}$ spectra at different bias currents and fixed $R_H$ corresponding to $n \approx 8 \times 10^{10}$ cm$^{-2}$. (d) $IR_H$ and $R_{2p}$ for 20 μA bias current. (e) $S_V^{1/2}$ and magnetic field noise $S_B^{1/2}$ at 1 kHz. (f) Bias current dependence of the minimum $S_B^{1/2}$ at 1 kHz. In panels d-f, error bars are determined from the standard error in the linear fit for $R_H$ and the standard deviation of points in the $S_V^{1/2}$ spectra in a window of width 200 Hz centered at 1 kHz.

12



\title{Supporting Information for:\\ Gate-Tunable Graphene Hall Sensors with High Magnetic Field Sensitivity}

\author{Brian~T.~Schaefer}\affiliation{Laboratory of Atomic and Solid-State Physics, Cornell University, Ithaca, NY 14853, USA}
\author{Lei~Wang}\affiliation{Kavli Institute at Cornell for Nanoscale Science, Cornell University, Ithaca, NY 14853, USA}
\author{Alexander~Jarjour}\affiliation{Laboratory of Atomic and Solid-State Physics, Cornell University, Ithaca, NY 14853, USA}
\author{Kenji~Watanabe}
\author{Takashi~Taniguchi}\affiliation{National Institute for Materials Science, 1-1 Namiki, Tsukuba 305-0044, Japan}
\author{Paul~L.~McEuen}
\author{Katja~C.~Nowack}\affiliation{Laboratory of Atomic and Solid-State Physics, Cornell University, Ithaca, NY 14853, USA}\affiliation{Kavli Institute at Cornell for Nanoscale Science, Cornell University, Ithaca, NY 14853, USA}\email{kcn34@cornell.edu}

\maketitle


\setcounter{figure}{0}    

\def\thefigure{S\arabic{figure}}
\def\theequation{S\arabic{equation}}
\def\thetable{S\arabic{table}}

\subsection{Device fabrication}

We obtain monolayer graphene (MLG), few-layer graphite (FLG), and \textapprox20-40 nm thick hexagonal boron nitride (hBN) flakes via mechanical exfoliation of bulk crystals using Scotch Magic tape onto as-received degenerately-doped silicon wafers with 285~nm SiO$_2$ (Nova Electronic Materials).
To increase the yield of large-area flakes, we clean the substrates with a gentle oxygen plasma, press the tape down onto the substrate, heat for 5 minutes at 100\degC{}, and let the chips return to room temperature before removing the tape~\cite{exfoliation}.
We were most successful using Kish graphite (CoorsTek) and hBN crystals grown using a high-pressure technique~\cite{hbn}.
We identify suitable flakes for devices only using optical inspection.

We create heterostructures with layer structure hBN/FLG/hBN/MLG/hBN/FLG/SiO$_2$/Si using a dry-transfer technique~\cite{lei, marcos, menyoungthesis}.
Our transfer slide consists of a thin sheet of poly(bisphenol A carbonate) (PC, Sigma Aldrich 435139) on top of a PDMS stamp (Gel-Pak) with curved top surface~\cite{hemisphere}, allowing for precise control over the engagement of the stamp onto the substrate.
The top hBN (\textapprox5~nm) only facilitates pickup of the other flakes and does not in principle influence the electronic properties of the device.
We pick up flakes sequentially at 80\degC{} and heat the final silicon substrate at 180\degC{} before releasing the stack, ensuring that bubbles trapped between the flakes are pushed towards the edges of the stack upon engaging~\cite{heterostructures, hotpickup}.
We intentionally misalign the straight edges of the graphene and hBN flakes by \textapprox$15^{\circ}$ to avoid creating a Moir\'e pattern between the graphene and hBN sheets~\cite{leigraphite}.
Finally, we dissolve the PC in chloroform for \textapprox4 hours, rinse with isopropyl alcohol, and blow dry with nitrogen.
A final anneal in high vacuum ($< 10^{-6}$~Torr) for 3 hours at 300\degC{} is effective in removing polymer residues from the transfer.

We employ standard nanofabrication techniques to pattern the device shape, expose a one-dimensional graphene edge~\cite{lei}, and make edge contacts (3~nm~Cr/40~nm~Pd/40~nm~Au or 3~nm~Cr/80~nm~Au) to the graphene and graphite layers.
Importantly, we have developed process conditions that help reduce the contact resistance. 
We use a CHF$_3$/O$_2$/Ar (20/10/10 sccm, 10 mTorr, 30 W ICP, 10 W RF) inductively-coupled plasma selective towards etching hBN.
Previous work suggests that selective etching reduces the contact resistance by increasing the metal-graphene contact area~\cite{moshe}.
Finally, we emphasize that to achieve consistently working contacts, we found it necessary to use an electron-beam evaporator with low base pressure (\textapprox$10^{-7}$~Torr) and a rotating sample chuck.

\clearpage
\subsection{Low charge inhomogeneity in graphite-gated devices}

\begin{figure*}[h]
\centering
\includegraphics[width=84.67mm]{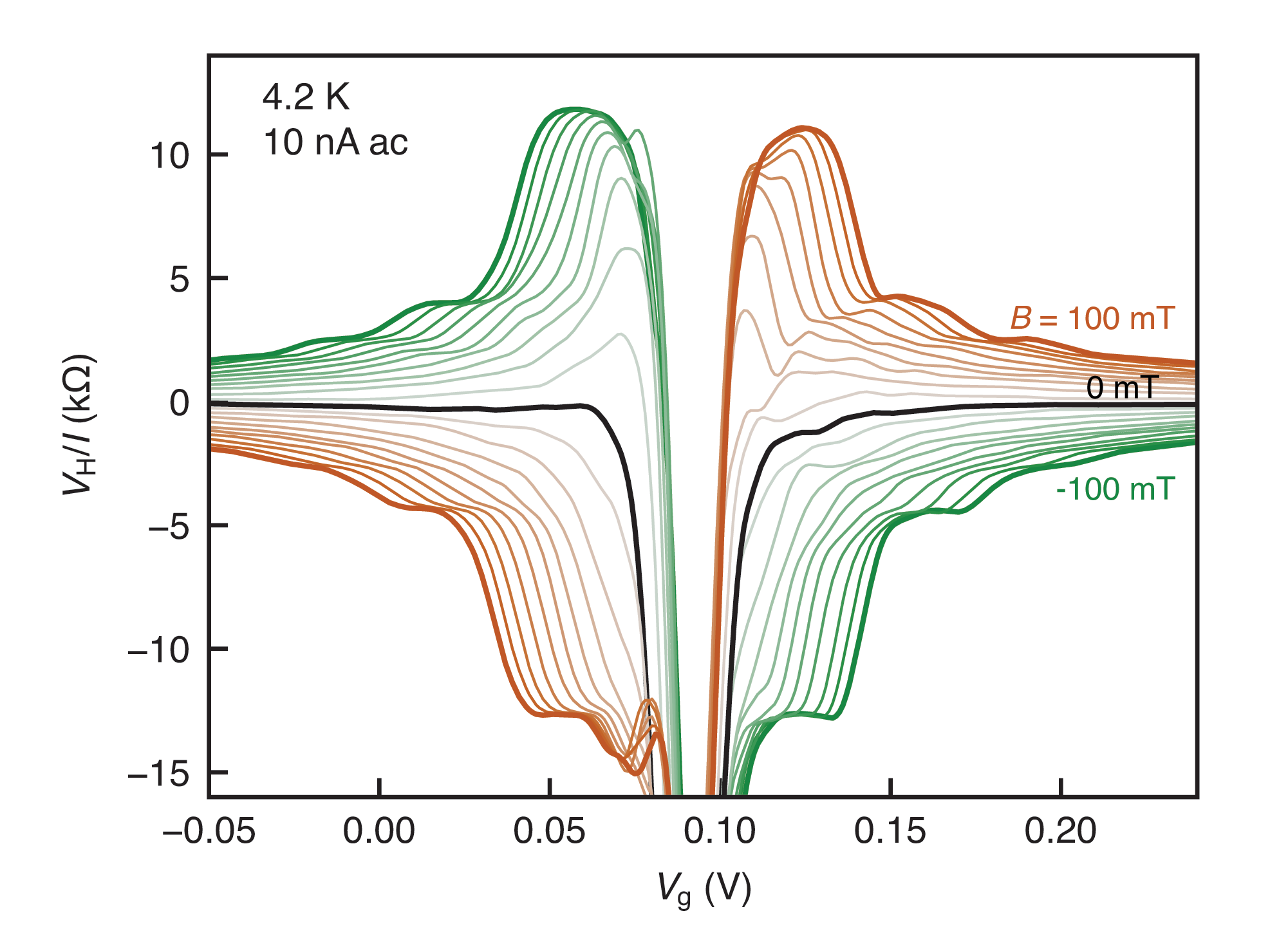}%
\caption{
Hall resistance $V_{\mathrm{H}}/I$ of device D1 measured versus gate voltage $V_{\mathrm{g}}$ and a series of magnetic fields $B$ in steps of 10 mT. 
\label{fig:Hallsupp}
}
\end{figure*}

The device discussed in the main text (D1) is fabricated with top and bottom graphite gate electrodes and possesses exceptionally small charge inhomogeneity $\delta n$.
In addition to the sharpness of the two-point resistance and large peak value of the Hall coefficient, the Hall resistance $V_{\mathrm{H}}/I$ exhibits quantum Hall resistance plateaus developing at magnetic field as low as \textapprox40~mT at liquid-helium temperature (Figure~\ref{fig:Hallsupp}).

\begin{figure*}[h]
\centering
\includegraphics[width=165.1mm]{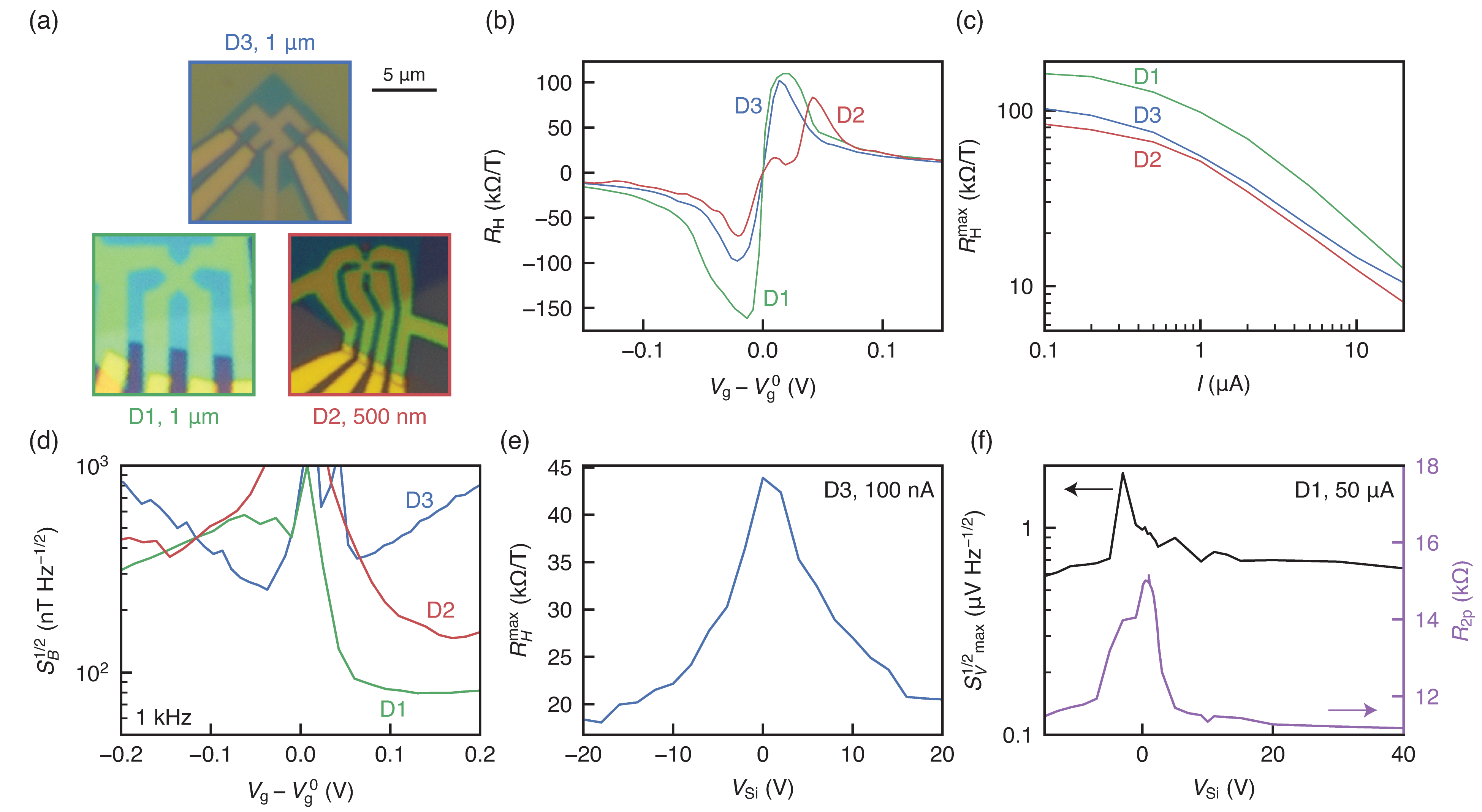}%
\caption{
(a) Optical images of three devices as described in Table~\ref{tab:devices}. 
(b) Hall coefficient ($R_{\mathrm{H}}$) measurements under 100~nA DC bias at liquid-helium temperature.
(c) Current bias dependence of peak $R_{\mathrm{H}}$.
(d) Magnetic field sensitivity $S_B^{1/2}$ at 1~kHz. We reach a minimum in $S_B^{1/2}$ for different DC current bias in each device: 20~\textmu{}A (D1), 5~\textmu{}A (D2), 2~\textmu{}A (D3). 
(e) Reduction in peak $R_{\mathrm{H}}$ upon applying voltage to the silicon gate of D3.
(f) Reduction of dc two-point resistance and peak voltage noise at 1~kHz upon applying silicon gate voltage to D1.
\label{fig:devices}
}
\end{figure*}

Table~\ref{tab:devices} and Figure~\ref{fig:devices} describe two additional devices: a 500~nm graphite-gated device (D2), and a 1~\textmu{}m device with a metal top gate (D3).
In D3, $\delta n$ is similar to that reported in silicon-gated hBN-encapsulated graphene devices~\cite{lei, hbninhomogeneity}. 
Although D1 and D2 possess the same layer structure, $\delta n$ in D2 is larger, which we speculate originates from poorly screened charge disorder from the device edges~\cite{transportreview, defects}.
The device size $w$ sets an approximate lower bound for the Fermi wavelength $\lambda_F=2\pi/\sqrt{\pi \delta n}\sim w$, giving $\delta n\sim5\times10^9$\densityunits for D2, in agreement with our measurements.

Other than lower $\delta n$, a second benefit of having a bottom graphite gate is that the contacts can be independently doped to high electron density while we gate the Hall cross to its most sensitive working point.
In D1, doping the contacts both reduces the two-point resistance and voltage noise (Figure~\ref{fig:devices}f) without decreasing the peak Hall coefficient.
However, gating D3 with the silicon gate significantly decreases the maximum Hall coefficient (Figure~\ref{fig:devices}e).

\begin{table*}[h]
\begin{tabular}{l || c | c |c|c|c}
                 				   & Size 	  & Bottom gate & Top gate                   	   & $\delta n$ (cm$^{-2}$\xspace)               & ${S_B^{1/2}}_{\textrm{min}}$ (nT\perroothz) \\	\hline
D1\footnote{from main text}	 & 1\micron  	 & FLG/hBN     & hBN/FLG                       					     & \textapprox$4\times10^9$  &        80     \\
D2               			  	 & 500~nm 	& FLG/hBN     & hBN/FLG                         				   & \textapprox$10^{10}$ &       150      \\
D3                 			 & 1\micron 	 & Si/SiO2/hBN & hBN/Ti/Au/Pt\footnote{5 nm Ti/30 nm Au/5 nm Pt} & \textapprox$10^{10}$  &            250
\end{tabular}\caption{Summary of additional devices\label{tab:devices}}
\end{table*}

\clearpage
\subsection{Carrier density gradient under large current bias}
Applying a large bias current to our devices strongly modifies the relationship between Hall coefficient and gate voltage.
Here, we show that our measurements are consistent with carrier density gradients resulting from the large bias current.

\begin{figure*}[h]
\centering
\includegraphics[width=165.1mm]{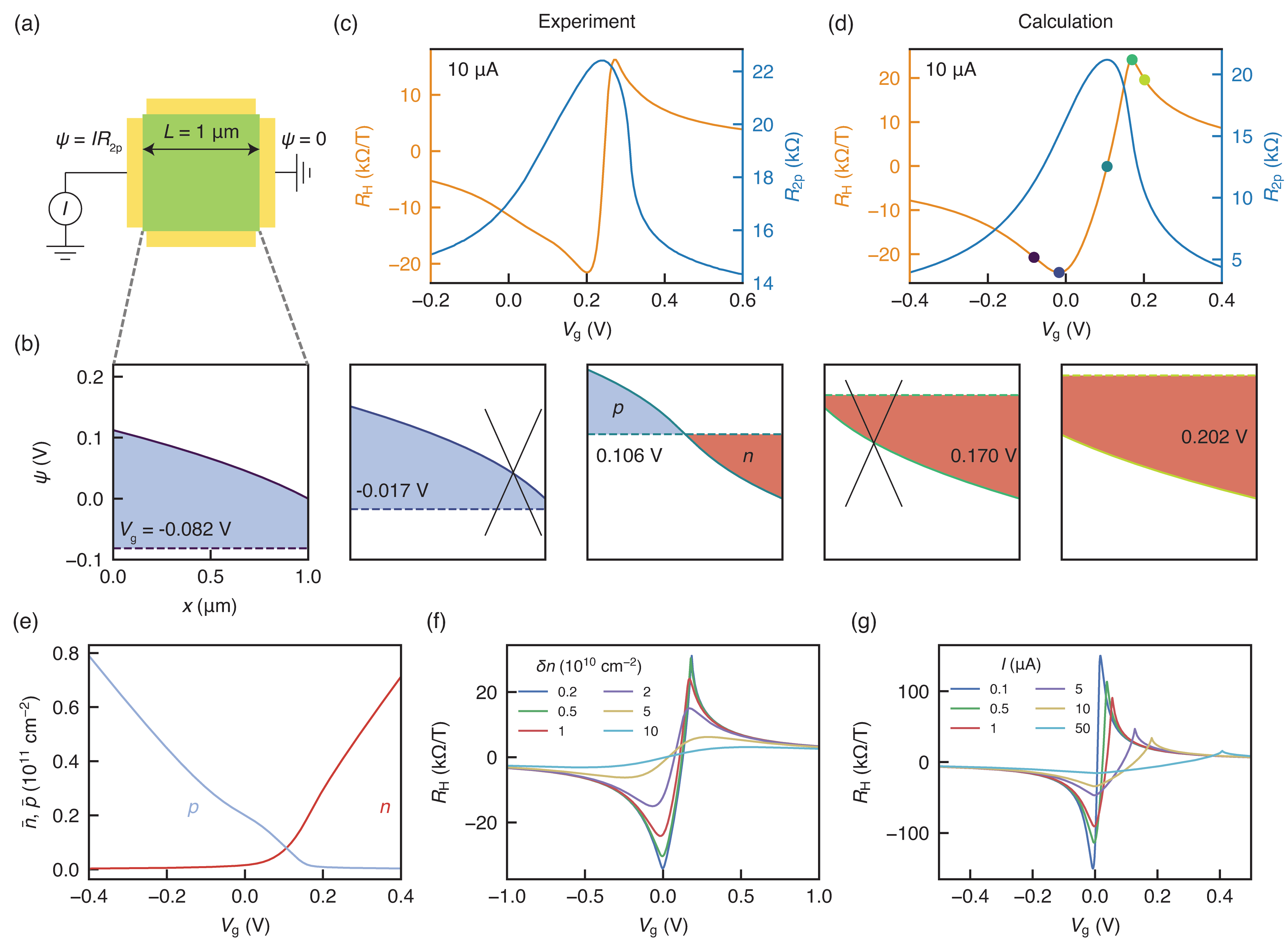}%
\caption{
(a) Schematic of the model device.
(b) Potential profiles across the Hall cross corresponding to the markers in (d). 
Dashed lines indicate the position of $V_g$, and the shading represents the carrier density (illustrated by Dirac cones in two of the panels).
(c,d) Measured (c) and calculated (d) $R_{\mathrm{H}}$ and $R_{\mathrm{2p}}$ under 10~\textmu{}A dc bias current. The calculation uses $\mu=20000$~cm$^2~$V$^{-1}~$s$^{-1}$, $C_g=0.03$\capacitanceunits, and $\delta n=10^{10}$\densityunits.
(e) Calculated average electron and hole densities in the Hall cross.
(f,g) Calculated charge inhomogeneity (f) and bias current (g) dependence of $R_{\mathrm{H}}$.
\label{fig:inhom}
}
\end{figure*}

We consider an $L\times L$ square device with contacts spanning the entire length of each of the four edges (Figure~\ref{fig:inhom}a) that measure the average Hall voltage in the center square.
The top and bottom contacts are the Hall voltage leads, the device is biased with constant current $I$ from the left contact (potential $\psi(x=0)=IR_{\text{2p}}$), and the right contact is grounded ($\psi(x=L)=0$).

The electron ($n_g$) and hole ($p_g$) densities away from the CNP depend on the potential difference between the gate and the graphene layer:
$$
n_g(x) = \frac{C_g}{e}[V_g-\psi(x)]
\hspace{1cm}
p_g(x) = \frac{C_g}{e}[\psi(x)-V_g],
$$
where $C_g$ is the gate capacitance.
Accounting for charge inhomogeneity $\delta n$ near the Dirac point, the electron and hole densities become~\cite{inhomogeneity}
$$
n(x) = \frac{n_g + \sqrt{n_g^2 + \delta n^2}}2\hspace{1cm}p(x) = \frac{p_g + \sqrt{p_g^2 + \delta n^2}}2.
$$
Noting $n_g^2=p_g^2$ and $n_g+p_g=0$, the total carrier density is:
$$
n(x)+p(x) = \sqrt{n_g^2 + \delta n^2} =  \sqrt{\frac{C_g^2}{e^2}[V_g-\psi(x)]^2 + \delta n^2}.
$$
Finally, using the resistivity $\rho^{-1} = e\mu(n+p)$, the Ohmic potential drop is given by:
$$
\frac{\partial \psi}{\partial x} = -\frac{I\rho(x)}{L}= -\frac{I}{Le\mu[n(x)+p(x)]}=  -\frac{I}{Le\mu\sqrt{\frac{C_g^2}{e^2}[V_g-\psi(x)]^2 + \delta n^2}}.
$$
Solving this differential equation numerically with initial condition $\psi(L)=0$ reveals that the potential $\psi(x)$ drops nonlinearly along the device channel (Figure~\ref{fig:inhom}b).
We extract the electron and hole densities $n(x)$ and $p(x)$ (Figure~\ref{fig:inhom}e), two-point resistance $R_{\text{2p}} = \psi(0)/I$, and average Hall coefficient $R_H$ using a two-carrier magnetoresistance model and average electron and hole densities in the channel~\cite{twocarriermodel}:
$$
R_H = \frac1e\frac{\bar{n}-\bar{p}}{(\bar{n}+\bar{p})^2}.
$$
Our calculation (Figure~\ref{fig:inhom}d) demonstrates many qualitative similarities to our measurements (Figure~\ref{fig:inhom}c), namely electron-hole asymmetry, a broadened Dirac peak and a reduced peak Hall coefficient.
Increasing the charge inhomogeneity (Figure~\ref{fig:inhom}f) or bias current (Figure~\ref{fig:inhom}g) further reduces the peak Hall coefficient, consistent with our measurements.

\clearpage
\subsection{DC transport and noise measurements}
The same wiring and instrumentation is used for both Hall voltage and noise measurements under dc current bias.
We apply dc current using a constant-current source and a series \textapprox1~M$\Omega$ bias resistor.
We amplify and filter the Hall voltage using a preamplifier (10 kHz lowpass filter), and we obtain time traces using the input terminal of a lock-in amplifier.
The preamplifier is in dc coupling mode for Hall voltage measurements and in ac coupling mode (with larger gain) for noise measurements.
In the latter case, we record 30 time traces sampled at 3.7~kHz for \textapprox4 seconds each, giving $2^8$ sampled points per time trace.
The Fourier transform of each time trace is computed using Welch's method~\cite{welch, numericalrecipes} with a Hann window.
We use frequency bins with 50\% overlap consisting of 27 points to reduce variance.
The resulting power spectral density $S_V$ is valid in a frequency band spanning \textapprox$1$~Hz to \textapprox3.7~kHz.
Noise levels quoted at a particular frequency are root-mean-square averages over a narrow band centered at that frequency, with the uncertainty given by the standard deviation of the data points in that band.

When the noise magnitude is above the noise floor of the instrumentation (input noise \textapprox6~nV\perroothz), the noise characteristics of the Hall voltage are well described by a combination of flicker (``$1/f$'') noise and random telegraph noise (RTN).
Meanwhile, the white Johnson noise $S_V^{1/2}=\sqrt{4k_BTR}$ is at most \textapprox$10$~nV\perroothz for a maximum $R_{\mathrm{2p}}$ of \textapprox$250$~k$\Omega$ at liquid-helium temperature (main text) or \textapprox$18$~nV\perroothz for \textapprox$20$~k$\Omega$ at room temperature (Figure~\ref{fig:temp}c).
In all cases, the Johnson noise is much smaller than the intrinsic charge noise measured in our devices.

\clearpage
\subsection{Random telegraph noise}
Although the general behavior of our devices remains the same between cooldowns, the specific amplitude of RTN and gate voltage region over which it is significant tend to change.
To illustrate this, we present noise measurements taken during two successive cooldowns, one in which RTN is only present for a small range of gate voltages and another in which RTN is almost completely absent.
These measurements are performed in the same way as in the main text, but the wiring used for these measurements involves twisted pairs which add a parasitic capacitance to ground that may suppress the noise slightly at frequencies approaching 1~kHz.

\begin{figure*}[h]
\centering
\includegraphics[width=165.1mm]{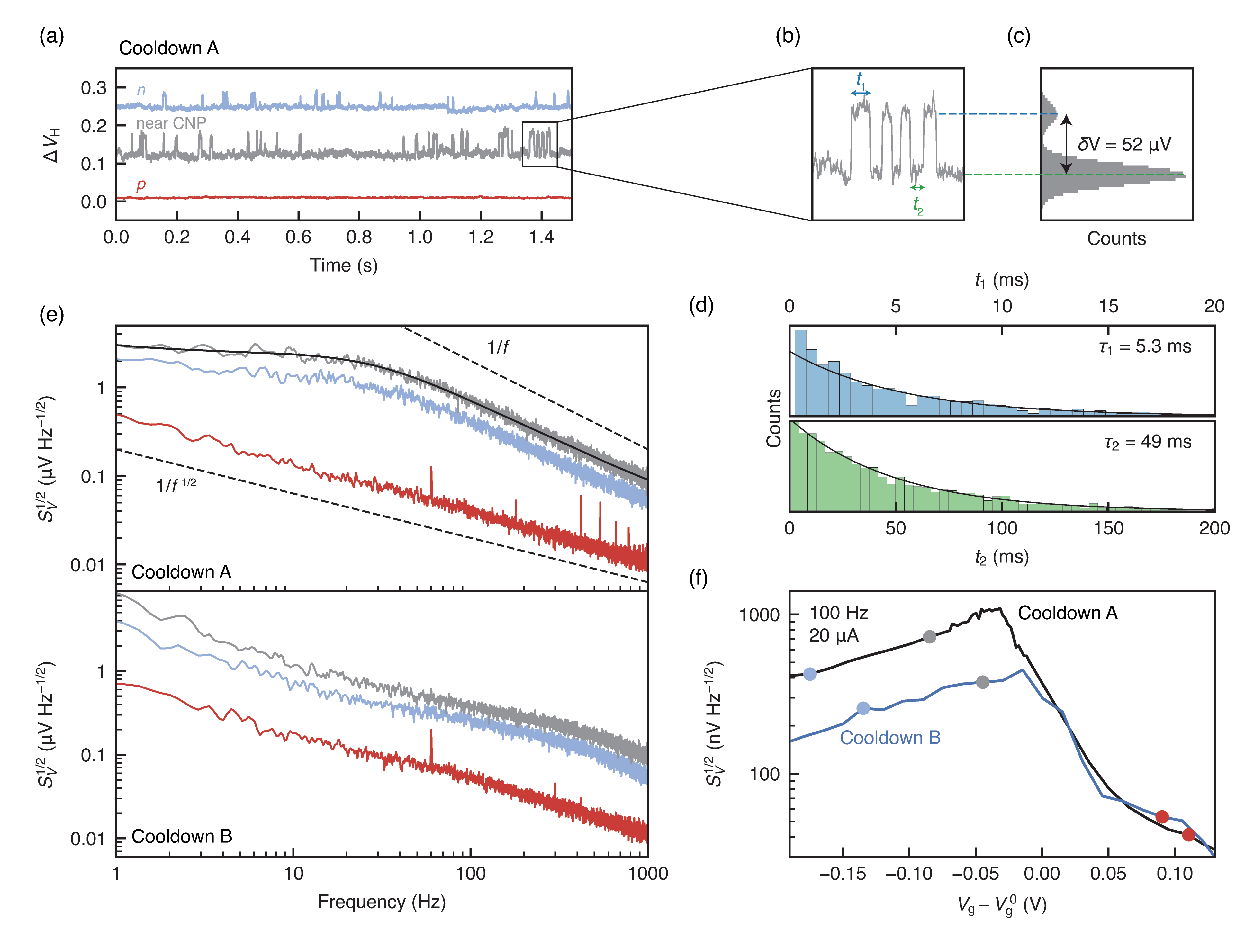}%
\caption{
(a) Hall voltage time traces at three different gate voltages, measured for device D1 during Cooldown A. Gate voltages correspond to the spectra in (e).
(b) Zoom-in of a voltage trace fluctuating between two voltage states with lifetimes $t_1$ and $t_2$.
(c) Voltage histogram from the entire 2.2-second time trace. 
(d) Histograms of the lifetimes of the two voltage states.
(e) $S_V^{1/2}$ spectra measured at 1~kHz. The solid curve is a fit to Equation~\ref{eq:rtn}. Both sets of spectra were acquired on device D1, but during separate cooldowns. Spectra correspond to the markers in (f).
(f) Average $S_V^{1/2}$ at 100~Hz.
\label{fig:rtn}
}
\end{figure*}

In Cooldown B (Figure~\ref{fig:rtn}e, lower panel), the nearly linear noise spectra are clearly dominated by 1/$f$-like noise, with a slight curvature due to weak RTN.
However, in Cooldown A (Figure~\ref{fig:rtn}e, upper panel), the noise spectra flatten below \textapprox30~Hz and fall off as $f^{-1}$ at high frequency, characteristic of a Lorentzian RTN spectrum~\cite{rtn}.
In the time domain, the voltage flucutates mainly between two distinct voltage states (Figure~\ref{fig:rtn}a,b).
The distribution of voltages comprising each of the two states is Gaussian (Figure~\ref{fig:rtn}c), while the lifetimes $t_1$ and $t_2$ each follow a Poisson distribution (Figure~\ref{fig:rtn}d)~\cite{rtntime}.
Fitting the lifetimes to an $\exp(-t/\tau)$ dependence yields a mean lifetime of $\tau_1= 3.9$~ms for the upper state and $\tau_2= 49$~ms for the lower state.

The total voltage noise spectral density can be modeled using \cite{rtn}
\begin{equation}
S_V = \frac{4\delta V^2}{\tau_1+\tau_2}\frac{\tau^2}{1+(2\pi f\tau)^2}+\frac{A}{f^\alpha},
\label{eq:rtn}
\end{equation}
where $f$ is the frequency, $\tau^{-1} = \tau_1^{-1}+\tau_2^{-1}$, $A$ is the flicker noise amplitude, and $\alpha\sim1$.
We fit the uppermost spectrum in Figure~\ref{fig:rtn}e (black curve) fixing $\alpha=1$ and obtain best-fit parameters $\delta V=52.5~\pm~0.5$~\textmu{}V, $\tau_1= 6.09~\pm~0.09$~ms, $\tau_2=49.0~\pm~0.9$~ms, and $A=(3.1~\pm~0.3)\times10^{-12}$~V.

\clearpage
\subsection{Temperature dependence of Hall coefficient and noise measurements at room temperature}
\begin{figure*}[h]
\centering
\includegraphics[width=165.1mm]{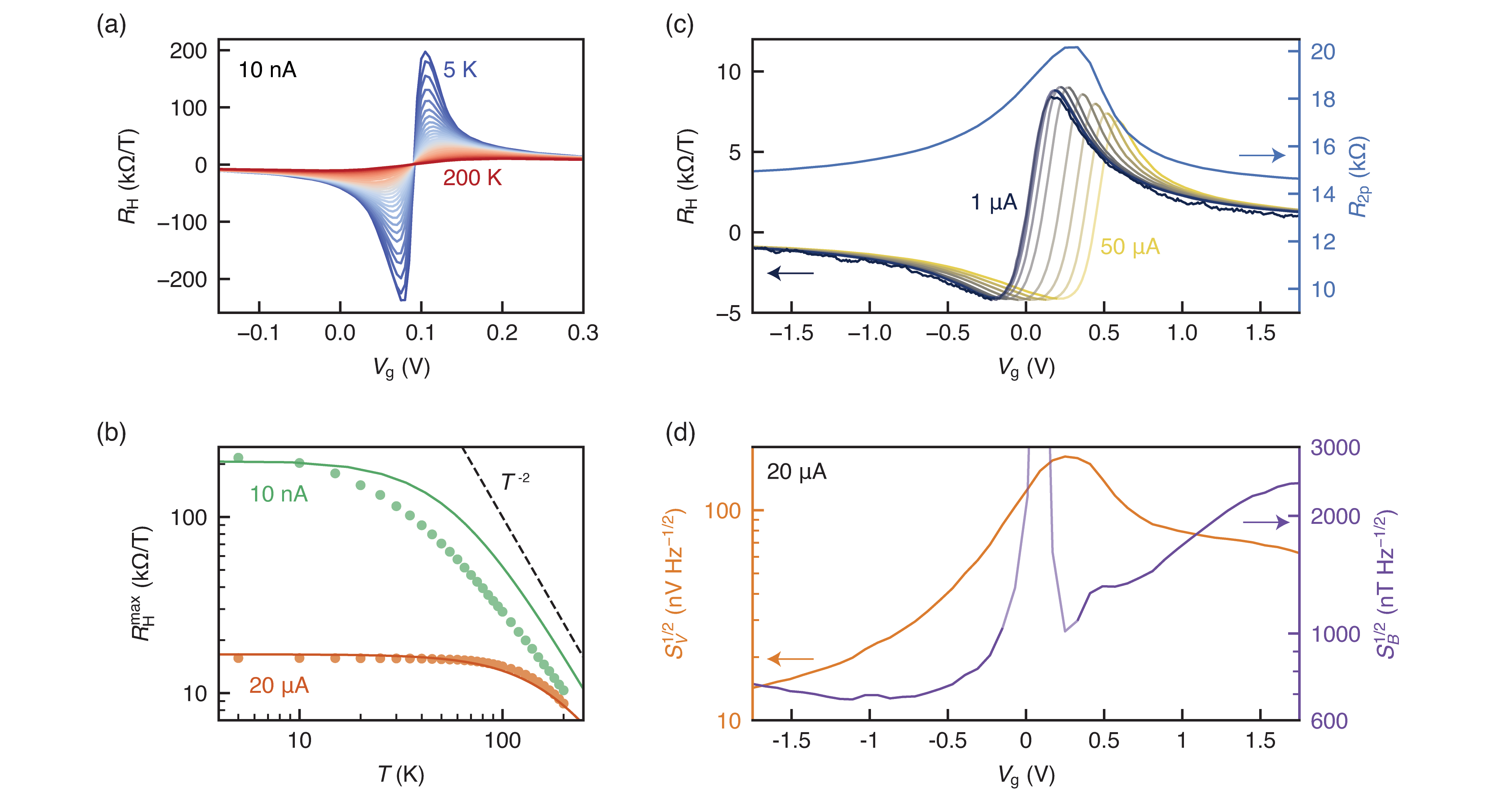}%
\caption{
(a) $R_{\mathrm{H}}$ measured as a function of temperature.
(b) Temperature dependence of peak $R_{\mathrm{H}}$ (markers) and comparison to the theoretical temperature dependence of charge inhomogeneity (solid curves).
(c) $R_{\mathrm{H}}$  and $R_{\mathrm{2p}}$ at room temperature for 20~\textmu{}A bias current.
(d) $S_{\mathrm{V}}^{1/2}$ and $S_{\mathrm{B}}^{1/2}$ at room temperature.
All measurements are performed on device D1.
\label{fig:temp}
}
\end{figure*}

Using a Quantum Design Physical Property Measurement System, we measure $R_{\mathrm{H}}$ as a function of gate voltage and temperature (Figure~\ref{fig:temp}a).
To save time, we estimate $R_{\mathrm{H}}$ using measurements only at $\pm50$~mT.
Extracting the peak $R_{\mathrm{H}}$ at each temperature (Figure~\ref{fig:temp}b), we observe that $R_{\mathrm{H}}^{\text{max}}$ shows weak temperature dependence at low temperature and decreases as $T^{-2}$ at high temperature.
Modeling the potential fluctuations due to charge disorder as a Gaussian distribution with amplitude $\Delta$, the charge inhomogeneity at the Dirac point is approximately~\cite{temptheory}
$$
\delta n(T) = \frac{1}{2\pi(\hbar v_{\text{F}})^2}\left[\Delta^2 + \frac{\pi^2}{3}(k_BT)^2\right],
$$
where $\hbar$ is the reduced Planck constant, $v_{\text{F}}=10^6$~m/s is the Fermi velocity, and $k_BT$ is the thermal energy.
In Figure~\ref{fig:temp}b, we plot $(\delta n(T)e)^{-1}$ for $\Delta=9$~meV (closely matching the 10~nA data) and $\Delta=32$~meV (closely matching the 20~\textmu{}A data).
For small bias, the crossover into the $T^{-2}$ regime occurs at a lower temperature than predicted by the model, likely due to reduction of $R_H$ via thermal activation of holes~\cite{temptheory2}.

At room temperature (\textapprox300~K), we perform full characterization of device D1 using the same cryostat insert used for low-temperature measurements, instead positioned between the poles of a C-frame electromagnet (GMW Associates, model 5403).
Notably, the bias current has little effect on $R_H$ below \textapprox$20$~\textmu{}A because the thermal charge inhomogeneity exceeds the additional effective inhomogeneity from the bias current (Figure~\ref{fig:temp}c).
Figure~\ref{fig:temp}d illustrates that $S_V^{1/2}$ and $S_B^{1/2}$ have a similar dependence on gate voltage as at low temperature, reaching a minimum $S_B^{1/2}\sim 700$~nT\perroothz for small hole doping.

\clearpage

\bibliographystyle{apsrev4-1-nospace}
\bibliography{GHP}

\end{document}